\documentclass[apjl]{emulateapj}

\shorttitle{PAH ionization versus fragmentation}
\shortauthors{Zhen et al.}

\begin{document}

\title{laboratory photo-chemistry of PAHs: ionization versus fragmentation }

\author{Junfeng Zhen$^{1,2\&3,4}$, Pablo Castellanos$^{1,2}$, Daniel M.\
  Paardekooper$^{2}$, Niels Ligterink$^{1,2}$, \\Harold Linnartz$^{2}$, Laurent Nahon$^{5}$, Christine Joblin$^{3,4}$, Alexander G. G. M.\
  Tielens$^{1}$} 

\affil{$^{1}$Leiden Observatory, University of Leiden, P.O.\ Box 9513, 2300 RA Leiden, The Netherlands} 
   
\affil{$^{2}$Sackler Laboratory for Astrophysics, Leiden Observatory, University of Leiden, P.O.\ Box 9513, 2300 RA Leiden, The Netherlands}

\affil{$^{3}$Universit\`e de Toulouse, UPS-OMP, IRAP, Toulouse, France}

\affil{$^{4}$CNRS, IRAP, 9 Av. colonel Roche, BP 44346, 31028, Toulouse Cedex 4, France}

\affil{$^{5}$Synchrotron SOLEIL, L$^{'}$Orme des Merisiers, 91192 Gif sur Yvette Cedex, France}
\email{Correspondence author: zhen@strw.leidenuniv.nl \\ Current address: Universit\`e de Toulouse, UPS-OMP, IRAP, Toulouse, France; CNRS, IRAP, 9 Av. colonel Roche, BP 44346, 31028, Toulouse Cedex 4, France. Electronic mail: junfeng.zhen@irap.omp.eu }

\begin{abstract}
Interstellar Polycyclic Aromatic Hydrocarbons (PAH) are expected to be strongly processed by Vacuum Ultra-Violet (VUV) photons. Here, we report experimental studies on the ionization and fragmentation of coronene (C$_{24}$H$_{12}$), ovalene (C$_{32}$H$_{14}$) and hexa-peri-hexabenzocoronene (HBC; C$_{42}$H$_{18}$) cations by exposure to synchrotron radiation in the range of $8-40$ eV. The results show that for small PAH cations such as coronene, fragmentation (H-loss) is more important than ionization. However, as the size increases, ionization becomes more and more important and for the HBC cation, ionization dominates. These results are discussed and it is concluded that, for large PAHs, fragmentation only becomes important when the photon energy has reached the highest ionization potential accessible. This implies that PAHs are even more photo-stable than previously thought. The implications of this experimental study for the photo-chemical evolution of PAHs in the interstellar medium (ISM) are briefly discussed.
\end{abstract}

\keywords{astrochemistry --- methods: laboratory --- ultraviolet: ISM --- ISM: molecules --- molecular processes}

\section{Introduction}
\label{sec:intro}

Strong emission features at 3.3, 6.2, 7.7, 8.6, 11.2 ~$\mu$m dominate the infraRed (IR) spectra of the ISM of the Milky Way as well as galaxies in the local and far Universe to redshift of $\sim$ 3 \citep{sel84,pug89,all89,gen98,armus07,lutz07,tie13}. These features are generally attributed to IR fluorescence of large ( $\sim$ 50 C atom) PAH molecules pumped by UV photons. These species must be abundant, ubiquitous, and contain $\sim$10\% of the elemental carbon and they play an important role in the ionization and energy balance of the ISM of galaxies \citep[and references therein]{tie08}. The astronomical PAH population is expected to exist in different charge and (de)hydrogenation states depending on the environment. As the IR spectral characteristics of PAHs are sensitive to the charge and (de)hydrogenation state, the observed spectrum of IR emission features will reflect the local physical conditions, and thereby offer probes of the environment of the emitting species \citep{tie05}. In the ISM, PAH molecules are necessarily exposed to continuous broadband UV radiation, and a large fraction of PAHs are expected to be singly, doubly, or even triply ionized. Indeed, models have predicted that for typical electron density and radiation field intensity values in photodissociation regions, the population of PAHs$^{++}$ may well exceed that of PAHs$^{+}$ \citep{bak01,bak02}. The possible presence of multiply-charged PAH ions in the ISM -- first proposed by  \citet{lea86} -- has received renewed interest with their potential involvement in the extended red emission \citep{wit06,mal07}.  As ionized PAHs are attractive candidates for the well-known but enigmatic diffuse interstellar bands, laboratory studies  have also focused on recording electronic spectra of PAH cations  \citep{salama11,steglich11}. The charge state of interstellar PAHs is of wider interest. UV photons can also lead to destruction of interstellar PAHs \citep{lep01,berne12}. Experiments have explored this fragmentation process \citep{eke98,jochims94,jochims96,zhe14,zhen2014} and the infrared characteristics of the resulting species have been quantum chemically calculated ( e.g. \citet{bau13,mac15}). Finally, detailed models have been constructed to assess their lifetime under different interstellar conditions \citep{lep01,mon13}.

\begin{figure*}[t]
  \centering
  \includegraphics[width=\textwidth]{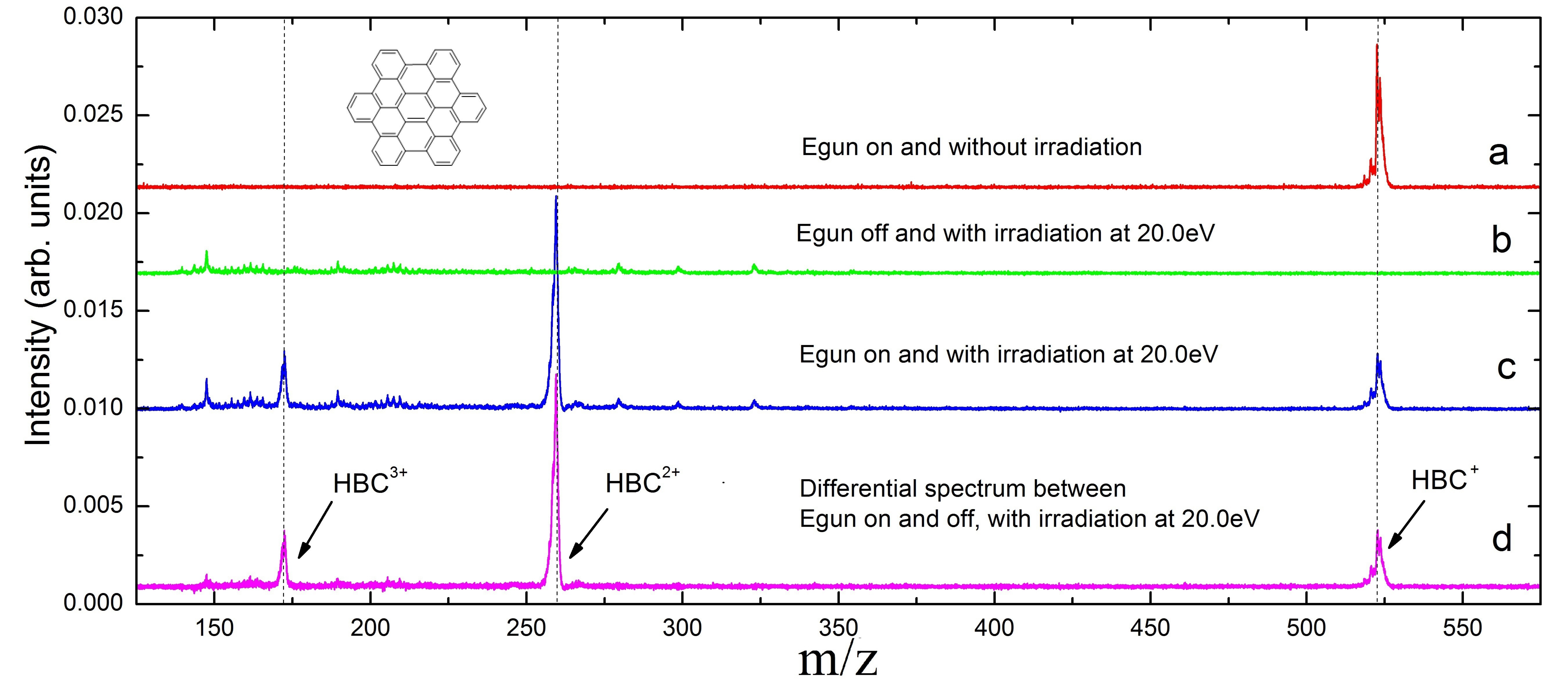}
  \caption{Mass spectrum of the products resulting from photo-irradiation of  HBC cations at 20.0eV (62.0nm): (a) Egun on and without irradiation; (b) Egun off and with irradiation. (c)  Egun on and with irradiation for 500 msec. (d) Differential mass spectrum between experiments (c) and (b).
  }
  \label{fig1}
\end{figure*}

\begin{figure}[t]
  \centering
  \includegraphics[width=\columnwidth]{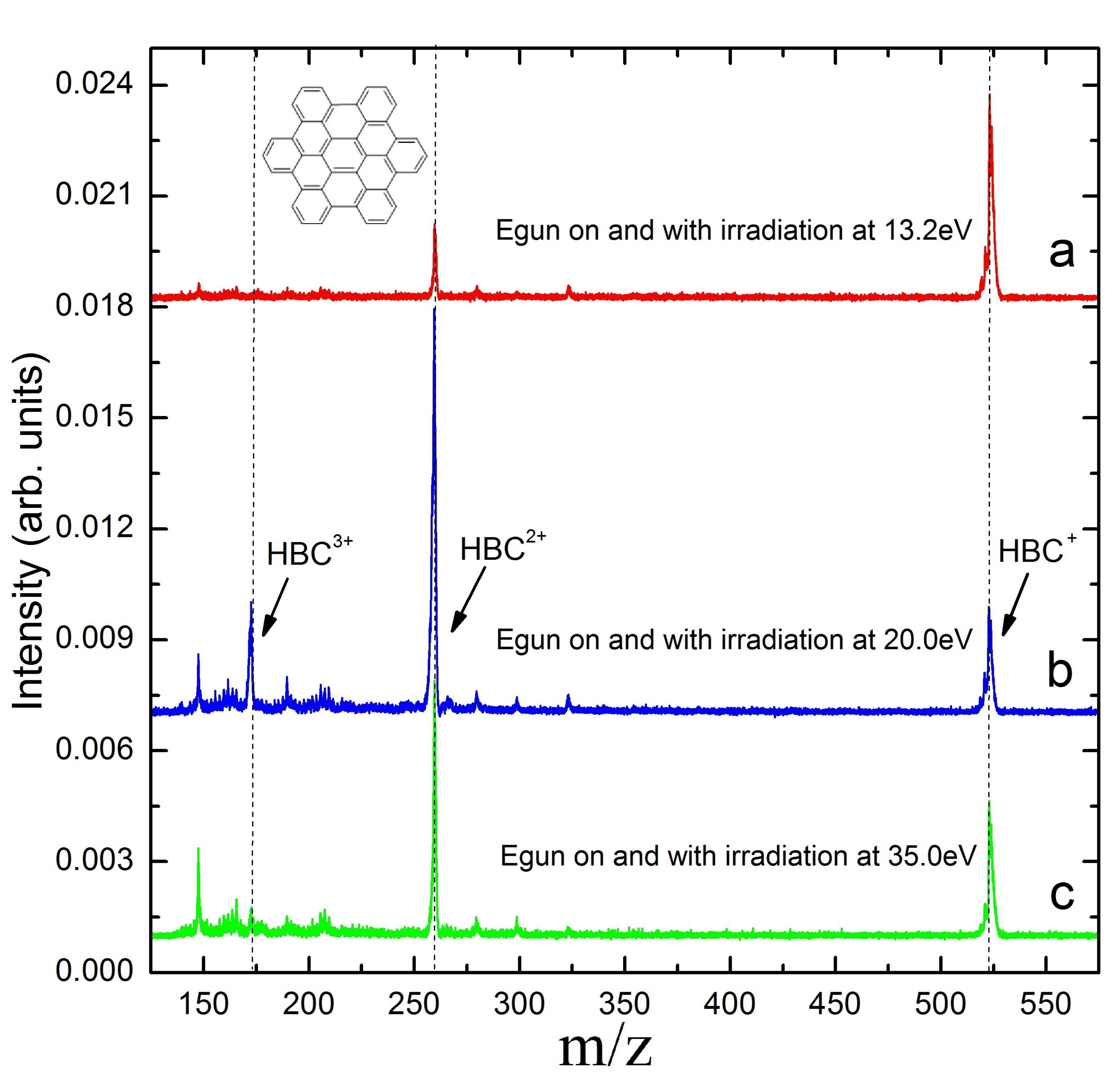}
  \caption{ Mass spectrum of the products resulting from photo-irradiation of HBC cations at different energies for 500 msec: (a) 13.2 eV (93.9nm), (b) 20.0eV (62.0nm) and (c) 35.0 eV (35.4nm).
  }
  \label{fig2}
\end{figure}

It is clear that interstellar PAHs are expected to be greatly influenced by the strong radiation field in space through ionization and fragmentation processes \citep{lea86,ver90,bak94,lep01,berne12,mon13}. However, until now, experimental studies have focused  either on the ionization or the fragmentation aspects \citep{jochims94,jochims96,jochims97,eke98,zhe14,zhen2014}. In fact, these two processes will be in competition and both aspects have to be considered simultaneously. In this letter, we perform a systematic study in the $8-40$ eV range of the photo-chemical behavior of three PAH cations that sample relevant sizes of interstellar PAHs using the SOLEIL synchrotron facility. The experiments are described in section~\ref{sec:exp}, the results are analysed in section~\ref{sec:results}, and discussed in section~\ref{sec:discussion}.

\section{Experimental Methods}
\label{sec:exp}

We have studied the photo-chemical processes of PAH cations in the $8-40$ eV range using i-PoP, our instrument for Photodissociation of PAHs connected to a VUV synchrotron beamline. i-PoP is described in detail in \citet{zhe14}. Briefly, PAHs are sublimated in an oven at an appropriate temperature, ionized by an electron gun (Egun), and then transported into a quadrupole ion-trap (QIT) via an ion gate. Once the ions are trapped, external electrical waveforms (SWIFT) are applied to the end cap electrodes to isolate a specific range of  mass/charge (m/z) species \citep{dor96}. After a short time delay, the ion cloud thermalizes to room temperature ($\sim$ 298K) through collisions with He buffer gas. The cations are then irradiated by synchrotron radiation. The ion-trap content is subsequently released and analyzed using a reflectron time-of-flight mass spectrometer. A LABVIEW program automates the full data acquisition process. 

The i-PoP setup is connected to the undulator-based VUV beamline DESIRS at the SOLEIL synchrotron facility in Saint-Aubin (France) \citep{nah12}. The DESIRS beamline is well suited for PAH cation studies because of its high photon flux, its broad and easy tunability in the 5-40 eV range, as well as its highly focused beam. In order to maximize the photon flux on our very diluted targets, we used the 6.65 m normal incidence monochromator at the zeroth order, leading to a typical photon flux in the 10$^{14}$ $-$ 10$^{15} $ photon s$^{-1}$ range in a 7 \% bandwidth. Another crucial feature of the beamline, in the context of mass spectrometry, lies in its spectral purity; i.e., the absence of any high harmonics of the undulator which are very efficiently filtered-out by a gas filter up to 16 eV (Argon) and by the cut-off of the grating coatings above 21eV.  Absolute incident photon fluxes are measured using calibrated photodiodes. The radiation beam is sent directly into the QIT, with a spot size of $\simeq $ 1.5 mm$^{2}$ in the center position. To achieve this, an additional chamber is connected with the differential pumping stage that accommodates the pressure difference between the i-PoP chamber (10$^{-7}$ mbar) and the beamline port (10$^{-8}$ mbar). The new chamber includes a retractable beam shutter (XRS2, Vincent Associates) that provides a minimum open time ( $\simeq $ 1 ms) under high-vacuum conditions. The beam shutter is externally triggered to guarantee that the ion cloud is irradiated for a specified amount of time during each scan cycle.

\begin{figure}[t]
  \centering
  \includegraphics[width=\columnwidth]{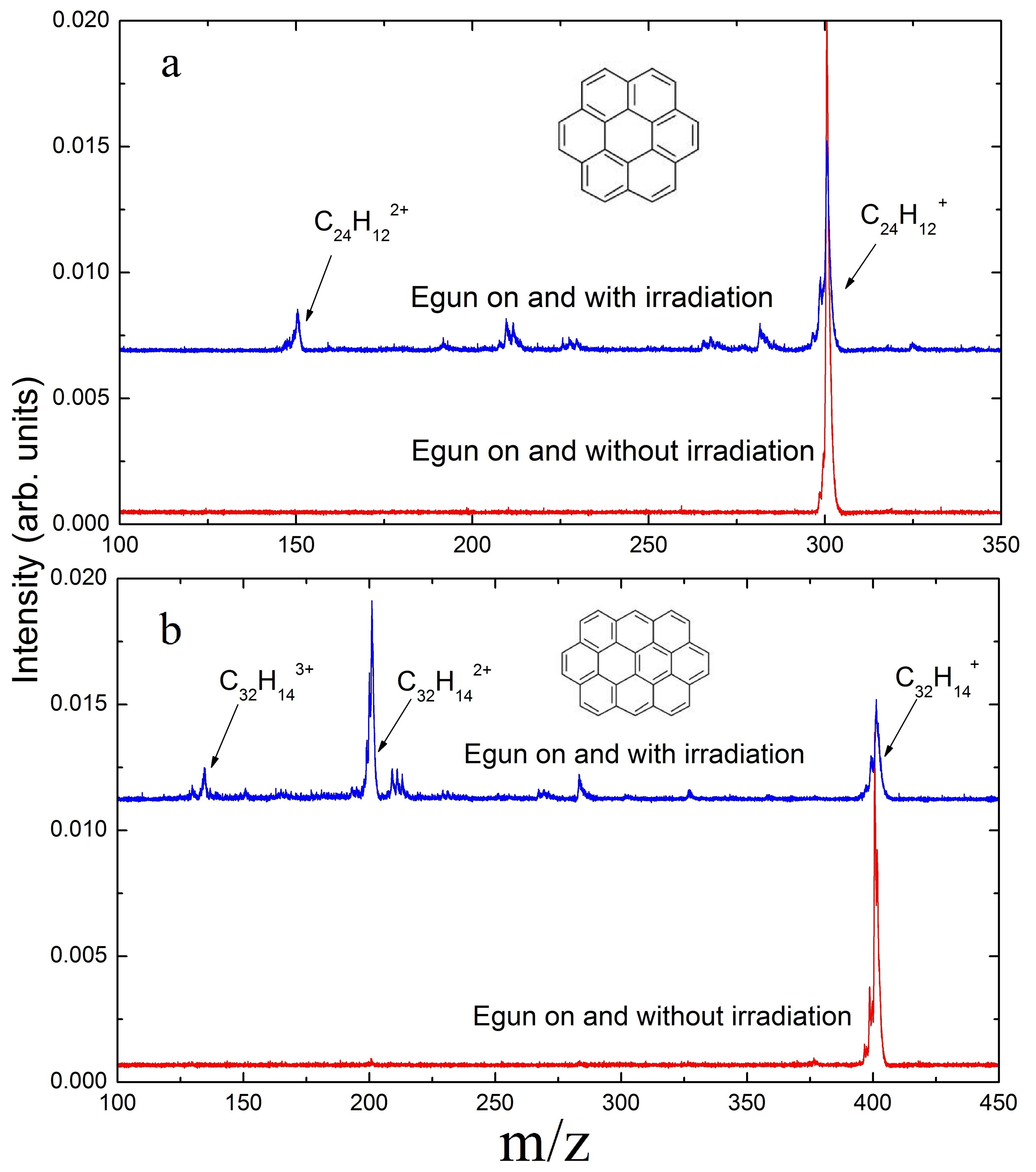}
  \caption{Upper Panel a: Mass spectrum of Corenene cation irradiated at 20.0eV (62.0nm) for 200ms, with and without irradiation. Lower Panel b: Mass spectrum of Ovalene cation irradiated at 21.0eV (59.0nm) for 500ms, with and without irradiation. 
  }
  \label{fig3}
\end{figure}  

\begin{figure*}[t]
  \centering
  \includegraphics[width=\textwidth]{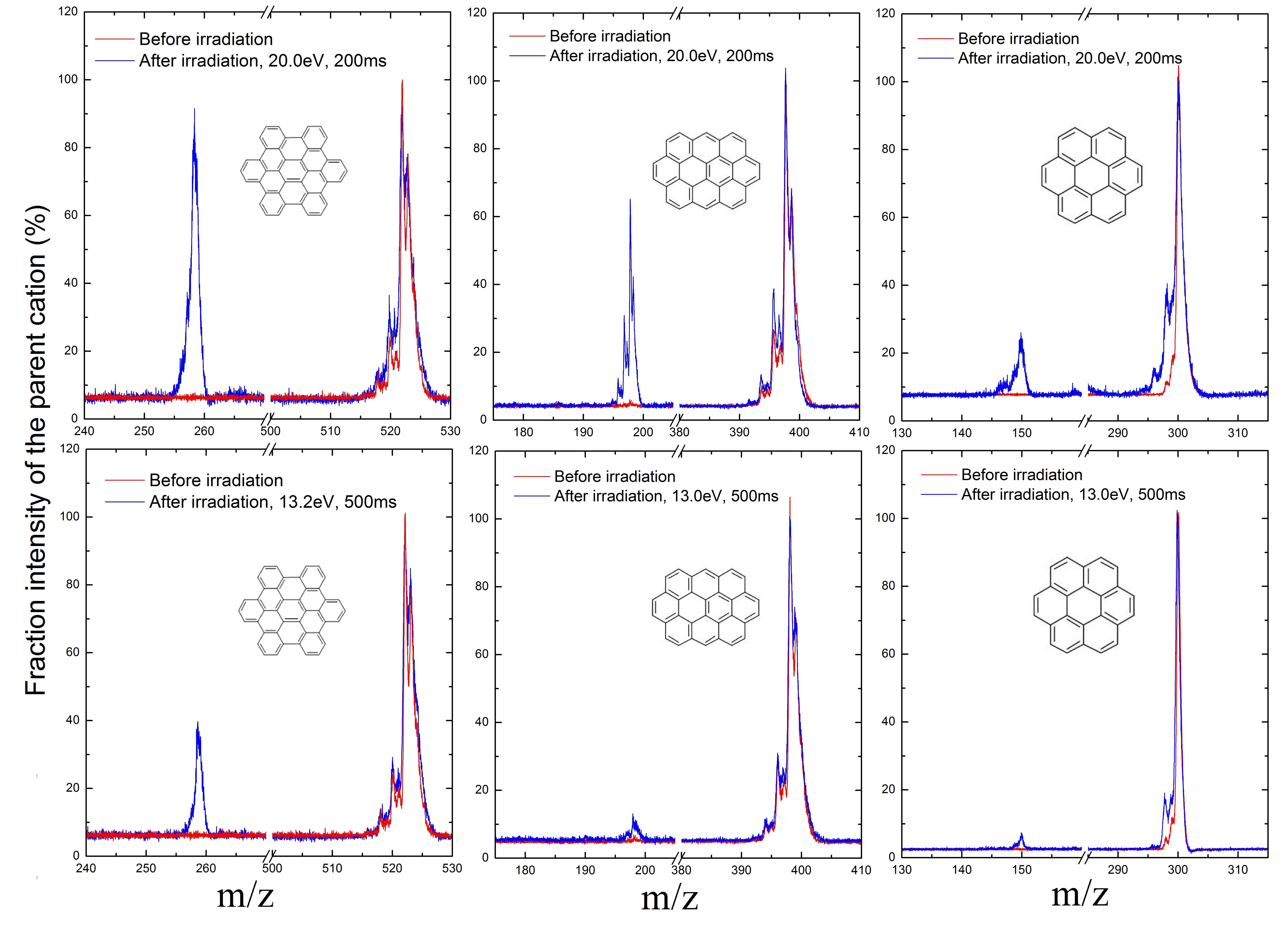}
  \caption{ Normalized intensity of PAH cations before and after irradiation: Top row, irradiation for 200 msec at 20.0eV (62.0nm). Bottom row irradiation for 500 msec at $\simeq 13.2$ eV (93.9nm). From left to right in each row, HBC, ovalene, and coronene cation. 
  }
  \label{fig4}
\end{figure*} 

\section{Results}
\label{sec:results}

We have studied the photo-chemical processing of three PAH cations: coronene (C$_{24}$H$_{12}^{\phantom{26}+}$; $m/z = 300.09$), ovalene (C$_{32}$H$_{14}^{\phantom{26}+}$; $m/z = 398.11$) and hexa-peri-hexabenzocoronene (C$_{42}$H$_{18}^{\phantom{26}+}$; HBC$^{+}$; $m/z = 522.14$). These particular PAHs are selected for their potential astrophysical interest \citep{tie05}, their commercial availability, their easy handling and, above all, because they span an astrophysically relevant range of sizes. The experimental results obtained with i-PoP are summarized in Figures $1-4$  and Table 1. 

Figure~\ref{fig1} summarizes the results for the HBC cation at 20.0eV (62.0nm). The mass spectrum before irradiation reveals a small amount of residual fragmentation (peaks due to 1H and 2H loss) as a byproduct of the electron impact ionization process as well as the presence of the $^{13}$C isotope (blended with the main mass peak) at natural abundance. After irradiation, a low level of contamination is present when no HBC is stored (Egun is off), due to the interaction of the synchrotron beam with PAH pollutions on the walls from previous experiments (Fig. 1b). 
These contaminating peaks are also present in the irradiation experiments of HBC$^+$ (Fig.~1c) but can be quite well corrected for (Fig.~1d). As the contamination peaks are sufficiently off-set from HBC$^+$ photo-products, they do not interfere with the analysis below, and we have not explicitly corrected the mass spectra in other experiments.

The HBC$^+$ irradiation results at 20.0 eV (Fig. 1d) show strong peaks due to the HBC dication (C$_{42}$H$_{18}^{\phantom{26}2+}$; HBC$^{2+}$; $m/z = 261.07$) and trication   (C$_{42}$H$_{18}^{\phantom{26}3+}$; HBC$^{3+}$; $m/z = 174.05$) but little evidence for loss of hydrogen (beyond that due to the effects of the electron gun; see below). Similar results are obtained at 35.0 eV (Fig.~2c), albeit that the overall photo-processing is less intense, reflecting the lower absorption cross section of HBC cation at this wavelength (35.4nm) compared to 62.0nm \citep{mal07}. At low photon energies (13.2eV; 93.9nm) (Fig.~2a), only the HBC dication is present. Fig. 2b is identical to Fig. 1c and is shown for ease of comparison for three different irradiation energies. Irradiation of ovalene shows grossly the same behavior with production of ovalene di- and tri-cations (C$_{32}$H$_{14}^{\phantom{26}2+}$; $m/z = 199.05$ and C$_{32}$H$_{14}^{\phantom{26}3+}$; $m/z = 132.67$), as illustrated in Figure~3b. We have selected to show ovalene at 21 eV with 500ms irradiation to bring out the small amount of C$_{32}$H$_{14}^{\phantom{26}3+}$ formation at these conditions. For coronene, the dication (C$_{24}$H$_{12}^{\phantom{26}2+}$; $m/z = 150.04$) is produced but the presence of the trication could not be confirmed as it is outside the QIT mass spectrometer's range (Fig.~3a) \citep{ray97}. Appearance energies for the different ion states are summarized in Table~1 and compared to measured, calculated, or estimated ionization potentials. We note that new ionization states become apparent immediately when the photon energy exceeds the ionization potential to the next ionization state. Hence, the appearance of the trication reflects the importance of (sequential) two photon processes. The importance of two photon processes is in agreement with estimates of the mean number of photons absorbed for these PAHs, given the large absorption cross section calculated by \citet{mal07}.

The fragmentation and ionization behavior of coronene, ovalene, and HBC cations at about 13.2 and 20eV are illustrated in Figure~\ref{fig4}. We note that for a given photon energy, the ionization yield increases with PAH size, i.e. from right to left, while the loss of H is more important for small PAHs. After 20 eV irradiation, the coronene and ovalene cations show clear evidence for the loss of two and four hydrogen atoms. We also note photo-fragments of the coronene and ovalene dications. With increasing energy, i.e., comparing the lower and corresponding upper panels, ionization yield and fragmentation yield increase but the same pattern remains: ionization completely dominates for HBC while fragmentation can compete with ionization for coronene. We can compare this quantitatively at high energies where ionization and fragmentation are expected to completely dominate over IR fluorescence. For fragmentation and ionization, we have ignored the small correction for the two photon process. We have corrected the strength of fragment peaks in the spectra for the background by subtracting a fraction of the e-gun fragmentation. To determine this fraction, we assume that fragments produced by the e-gun will be ionized at the same way as the parent and isotopic species. Hence, we ignore the effect of small shift in ionization potential expected for the fragments \citep{mal2008}. With this assumption, the fraction is then determined by scaling the strength of the parent plus isotope in the spectrum by the parent plus isotope in the e-gun spectrum.
In this way, we estimate that -- at 20eV -- only around $\sim$ 3\% of the HBC cations fragment while $\sim$ 97\% are ionized. For ovalene and coronene, the ionization yields are 70 $\pm$ 10\%\ and 25 $\pm$ 3\%, respectively (e.g.,  fragmentation yields of 30 $\pm$ 10\%\ and 75 $\pm$ 3\%). Details of the procedures will be reported in a forthcoming paper  \citep{castellanos15}. 

\begin{table*}
\caption{Appearance energies and ionization potentials for doubly- and triply- charged PAH parent ions. \label{BenchCalcs}}\small
\begin{tabular}{ccccccc}
\hline
Name & Formula & \multicolumn{2}{c}{Ionization potential$^a$} &\multicolumn{2}{c}{Appearance potential$^a $} \\ 
 &&PAH$^{+}$ $\to$ PAH$^{2+}$& PAH$^{2+}$ $\to$ PAH$^{3+}$&PAH$^{+}$ $\to$ PAH$^{2+}$& PAH$^{2+}$ $\to$ PAH$^{3+}$\\
\hline
hexa-peri-hexabenzocoronene \\(HBC)  & C$_{42}$H$_{18}^{\phantom{22}+}$ &9.0$^e$ &  13.0$^c$  & $\leq$ 8.5$\pm$ 0.6 $^f$ & 16.5$\pm$ 2.0\\
ovalene  & C$_{32}$H$_{14}^{\phantom{22}+}$& 9.6$^e$&15.4$^c$ & 9.4$\pm$ 0.6 & 18.0$\pm$ 2.0\\
coronene & C$_{24}$H$_{12}^{\phantom{22}+}$& 11.3$^b$&17.2$^c$& 10.5$\pm$ 0.7 & $-^d$ \\
\hline
\end{tabular}
\\
{$^a$} In eV;
$^b$ \citep{tob94};\\
$^c$ Estimated from circular disk approximation \citep{tie05};
$^d$ Outside of the QIT mass spectrometer range;\\
$^e$ \citep{mal2007};
$^f$ The HBC dication is already present at the lowest energy in our experiments.
\end{table*}

\section{Discussion and astronomical implications}
\label{sec:discussion}

At the high photon energies of our experiments, several processes -- including ionization and fragmentation -- can compete in the relaxation process. Calculated UV absorption cross sections for neutral and ionized PAHs reveal broad, blended peaks due to $\pi^{\star}\leftarrow\pi$, $\pi^{\star}\leftarrow\sigma$, $\sigma^{\star}\leftarrow\pi$, and $\sigma^{\star}\leftarrow\sigma$ transitions, involving superexcited (bound) states, which lie above the ionization continuum \citep{mal07}. Apart from these single electron transitions, valence shell two-electron processes in which more than one electron is involved, may also be important \citep{becker96}, particularly at higher energies ($\gtrsim16 $ eV; \cite{mar15}). Due to the high electron density of states, coupling is very fast among these states, and highly excited PAHs can rapidly (tens of femtoseconds; \cite{mar15}) relax to lower lying states -- through conical intersections -- including the ionizing continuum. If no ionization occurs, this internal conversion will leave the species in a highly vibrationally excited state in the ground electronic state. Internal vibrational relaxation will then occur, sharing the vibrational energy among all the modes on a timescale of tens of nanoseconds. The energy will eventually be radiated away in the IR through vibrational transitions on a timescale of $\sim 1$ sec, but if the excitation is sufficiently high, enough energy can accumulate in one vibrational mode corresponding to a given C-H bond, leading to the unimolecular fragmentation of the cation in the electronic ground state. Ionization competes, thus, with Internal Conversion, which is then followed by Internal Vibrational Relaxation (IVR), and then fragmentation or IR emission. Experiments on small, neutral PAHs (up to coronene) have shown that the ionization yield increases with increased photon energy and typically reaches unity some 9 eV above the ionization potential \citep{jochims96,jochims97}. For coronene cations, experiments show that fragmentation (H-loss) occurs on a timescale $10^{-4}$ sec for an internal energy of $\simeq 12$ eV  \citep{jochims94}. In the ISM, fragmentation can occur on longer timescales at which fragmentation is in competition with IR cooling. This happens for coronene cation at an internal energy of $\sim 9$ eV  \citep{mon13}. 

In the analysis of the experimental results presented in section~\ref{sec:results}, several points have to be kept in mind. First, for all PAHs, the UV absorption cross section per C-atom is very similar in strength and in general behavior with energy; e.g., the absolute cross section is almost a factor 2 larger for HBC (42 C-atoms)  than for coronene (24 C-atoms).  Second, the UV absorption cross section typically increases with photon energy by a factor 10 from $\simeq 8$ eV to a broad ($\simeq 2$ eV) maximum around $\simeq 16$ eV and then decreases again by a factor 3 to 25 eV \citep{mal07}. Third, the synchrotron beam may not overlap with the ion cloud in the same way for the different species. All these points imply that we should only consider the relative importance of the different processes when comparing different species or when comparing measurements for the same species at different energies. 

The experimental results reveal that ionization competes well with fragmentation for PAHs  cations as small as coronene but for PAHs as large as HBC ionization dominates. We consider that two effects play a role. First, the larger number of electrons lead to a higher density of electronic states. Hence, the electronically excited species have to traverse more ``electronic'' phase space and drop through more conical intersections before reaching the ground electronic state. The associated longer ``electronic'' lifetime facilitates transfer to the ionizing continuum and will lead to a relatively higher ionization yield.  Second, a larger species will need a higher internal energy to reach the same level of excitation because of the higher number of vibrational modes into which the energy can be distributed. Specifically, for HBC, the internal energy has to be almost a factor 2 higher than for coronene for the same excitation level of the vibrational modes. Hence, the fragmentation yield will rapidly decrease when going from coronene to HBC \citep{castellanos15}. While, we have focused here on the behavior of HBC$^+$, we note that HBC$^{2+}$ shows a similar behavior with little \textbf{apparent} fragmentation compared to ionization over the relevant energy range. 

Our results imply that large PAH cations are inherently photo-stable in the ISM. Fragmentation will become important when only electronic states below the next ionization potential of the species are accessible, otherwise photoionization will quench the competing photodissociation process. For HBC in HI region (with $h\nu\leq 13.6$ eV), this corresponds to HBC$^{2+}$. For larger species such as circumcoronene (C$_{54}$H$_{18}$) even the tri-cation is accessible with photons less than 13.6 eV. Fragmentation is then not expected to become important until the ionization parameter, $\gamma$ ($=G_oT^{1/2}/n_e$ with $G_o$ the strength of the radiation field in units of the Habing field, $T$ the gas temperature, and $n_e$ the electron density), exceeds $\simeq 10^6$ \citep{tie05} and, thus, only close to a bright star. Some fragmentation of HBC$^+$ may, in principle, occur when HBC$^{2+}$ recombines. However, recombination limits the internal excitation energy of HBC$^+$ to the ionization potential (9 eV for HBC$^+$) and little fragmentation will occur at such low energies for large species \citep{mon13,castellanos15}.

In the reflection nebula, NGC 7023, observations have revealed that PAHs are rapidly destroyed near the illuminating star while the abundance of the fullerene, C$_{60}$ increases \citep{berne12}. A similar behavior is observed in other regions in space \citep{castellanos14}. This has been interpreted as the result of photochemical evolution leading to loss of  H-atoms, the formation of graphene sheets, followed by a curling up of these sheets into cages and then fullerenes \citep{berne12, mon13, berne15}. There is good experimental support for the process of H-loss and the formation of pure carbon sheets, clusters, or cages \citep{eke98,zhe14,west15}. Recently, we have also obtained experimental support for the formation of fullerenes through this process \citep{zhen2014}. These latter studies were specifically designed to avoid the ionization versus fragmentation competition by pumping the molecules through the sequential events of multi-photon absorptions all below the ionization potential. In conjunction with the study reported here, we infer that H-loss, graphene formation, and cage \&\ fullerene formation will only occur when ionization is a marginal process. As these experiments demonstrate, multi-photon events are one, very relevant scenario and the theoretical study by \cite{mon13} has already pointed out the importance of multi-photon events in the fragmention of interstellar PAHs in neutral hydrogen regions in the ISM. These absorption events should then involve photons below the ionization potential of the species involved. Alternatively, in HII regions, much higher energy photons are available and, in principle, ionization and fragmentation by a single photon may become possible. However, we note that only a small amount ($\leq2$ \% ) of fragmentation of HBC$^+$ even for photon energies as high as 35.0 eV, some 26.0 eV above the ionization potential.

\section{Conclusion}
\label{sec:concl}

Photo-chemical processing with synchrotron irradiation of PAH cations in the 
range of $8-40$ eV reveals that ionization becomes more and more 
important when the size of the PAH increases, quenching the competing photodissociation process. For PAHs of some 
50 C-atoms, ionization fully dominates over fragmentation. 
Hence, in addition to the kinetic parameters involved in the 
fragmentation process, the ionization process has to be quantified 
for a proper assessment of the stability of PAHs in space. 
In other words, we consider it likely that variations 
in the ionization characteristics of PAHs are as important in evaluating  the composition of the interstellar 
PAH family as the kinetic fragmentation parameters. The ionization 
state may well leave its imprint on the IR spectrum of PAHs \citep{mal07,bak01} and, for example, ionized large PAHs have been proposed as the carriers of a characteristic band at 7.90 $\mu$m prominent in some planetary nebula \citep{joblin08}.  Further observational studies -- supported by experimental or 
quantum chemical studies -- are warranted. 

\acknowledgments

We are grateful to M. J. A.\ Witlox and R.\ Koehler for technical
support. Studies of interstellar chemistry at Leiden Observatory
are supported through advanced-ERC grant 246976 from the European
Research Council, through a grant by the Dutch Science Agency, NWO, as
part of the Dutch Astrochemistry Network, and through the Spinoza
premie from the Dutch Science Agency, NWO. We also acknowledge 
support from the European Research Council under the European Union's Seventh Framework
Programme ERC-2013-SyG, Grant Agreement n.~610256 NANOCOSMOS, as well as from the EU Transnational Access Program CALYPSO. We are indebted to J.-F. Gil for its technical help in installing the i-PoP set-up onto the DESIRS beamline, and to the general staff of SOLEIL for running the facility under project n.~20130911.

\end{document}